# MIDDLE EAST TECHNICAL UNIVERSITY
# DEPARTMENT OF STATISTICS

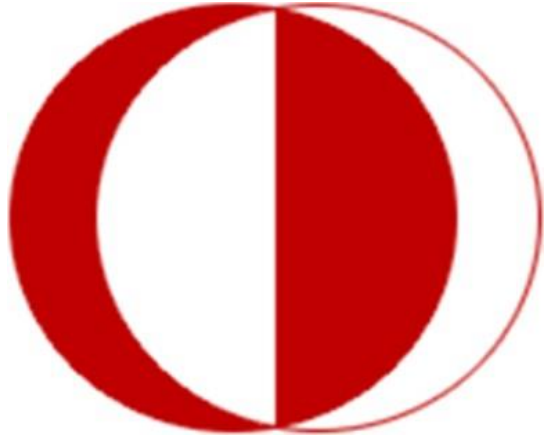

# METU STUDENTS' COLLEGE LIFE SATISFACTION

Furkan Berk DANIŞMAN

Gizem SARUL

Niyousha AMİNİ

Osman Orçun ADA

Sena Gülizar AKTAŞ

Sıla İlyürek KILIÇ

**ANKARA**

**2022-2023**

# Abstract

The research was conducted to identify the factors that influence college students' satisfaction with their college experience. Firstly, the study was focused on the literature review to determine relevant factors that have been previously studied in the literature. Then, the survey analysis examined three main independent factors that have been found to be related to college students' satisfaction: Major Satisfaction, Social Self-Efficacy, and Academic Performance. The findings of the study suggested that the most important factor affecting students' satisfaction with their college experience is their satisfaction with their chosen major. This means that students who are satisfied with the major they have chosen are more likely to be overall satisfied with their college experience. It's worth noting that, while the study found that major satisfaction is the most crucial factor, it doesn't mean that other factors such as Social Self-Efficacy, Academic Performance, and Campus Life Satisfaction are not important. Based on these findings, it is recommend that students prioritize their major satisfaction when making college choices in order to maximize their overall satisfaction with their college experience.

# TABLE OF CONTENT

# 1-) Introduction

The study aims to explore whether and how major satisfaction, social self-efficacy, and academic performance are associated with METU students' satisfaction with campus life. Besides using major satisfaction, social self-efficacy, and academic performance as quantitative and qualitative indicators, the residential place of the students is used as a categorical variable as well, since the location of the METU campus might be considered as a "non-city-center" location. In addition, the literature indicates that not only there might be a relationship between these factors to the campus life satisfaction of students, but also there is a significant relationship between the factors as well. Therefore, prior research and investigations are strongly supporting the idea that there is a correlation between the dependent variable and affecting factors as much as the correlation between the factors. In light of this potential, the research seeks to give a solution and provide suggestions to those METU students who are dissatisfied and meet difficulties to achieve their goals and dreams during their college years. The following diagram illustrates the connection between the identified factors and campus life satisfaction.

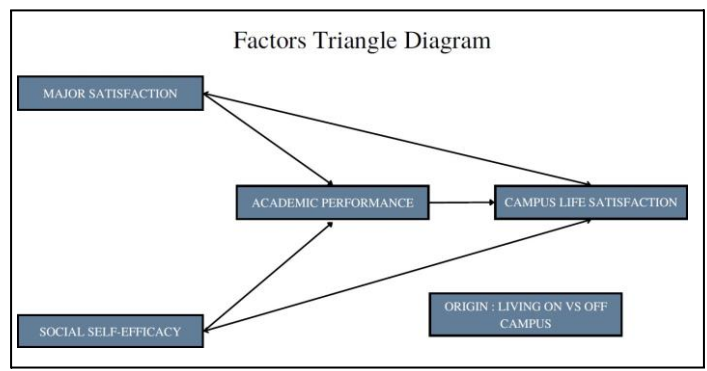

*Figure 1: Connection between the factors and campus life satisfaction*

## 1.1-) Aim of the research

This research aims to identify the key factors that influence student satisfaction with their experience on a college campus. These factors can include major satisfaction, social self-efficacy, and academic performance. The goal is to use statistical models to better understand the relationships between these factors and student satisfaction, and to use this information to guide future research studies. This research will help in understanding the factors that promote student satisfaction and can help the college administration to improve the student's life on campus.

## 1.2-) Research Questions

In this research, there are three factors that affect students's campus life satisfaction - major satisfaction, self efficacy, academic performance. Research questions were illustrated in the following table.

| Attribute | Null Hypothesis ($H_0$) | Alternative Hypothesis ($H_1$) |
|---|---|---|
| Major Satisfaction | There is no statistically significant difference between means of male and female students' overall major satisfaction. | There is a statistically significant difference between means of male and female students' overall major satisfaction. |
| Social Self Efficiency | Students with a 3.5 or higher social self efficacy value produce equal and lower academic performance than Students with a lower social self efficacy value. | Students with a 3.5 or higher social self efficacy value produce higher academic performance than Students with a lower social self efficacy value. |
| Academic Performance | There is no statistically difference in the average campus life satisfaction between students with an average of over 3.0 GPA and students with an average of less than 3.0 GPA. | There is statistically difference in the average campus life satisfaction between students with an average of over 3.0 GPA and students with an average of less than 3.0 GPA. |

*Table 1: Statistical hypothesis and factors*

## 1.3-) Survey Description

The survey of college life satisfaction among METU students was conducted, consisting of 57 questions covering various aspects of their university experience such as satisfaction with their major, academic performance, and social self-efficacy. The survey utilized a variety of question formats including Likert scales, drop-down options, open-ended questions, and matrix questions. It aimed to gather both quantitative and qualitative data to gain an understanding of the students' overall satisfaction with their college experience. 249 students participated in the survey, which was administered through a combination of online and in-person methods, with a response rate of 80%.

## 2-) Review of Literature

### 2.1-) Literature Review on Major Satisfaction

The studies on how to feel well or at least better are best conceptualized under the name "subjective well-being" (Diener, 1999). Diener defines subjective well-being as "people's evaluations of their lives—evaluations that are both affective and cognitive" (Diener, 2000). Pesch et al. (2018) indicated in the research new concept is defined as ''enjoyment of one's role or experiences as a student'' (Lent et al., 2007) to evaluate college students' satisfaction. Among all the forms of academic satisfaction, major satisfaction has received the most attention in the experimental literature.". Students' satisfaction with their academic major may predict academic performance, social self-efficacy, and overall life satisfaction.

Kim Hee-Yung (2009) and Pesch et al. (2018) also made research regarding major Satisfaction. They focused on how to scale major satisfaction and which factors have an effect on it. However, in this research, the focus is to advance the progress that has been made and measure major satisfaction by using these findings and analyzing its psychometric properties to check whether there is an association between major satisfaction and the campus life satisfaction of METU students.

### 2.2-) Literature Review on Social Self-Efficacy

Self-efficacy has been defined as a person's belief in successfully performing a task (Bandura, 1997). Having this competence in the social domain allows a person to be effective in social interactions and to establish positive interpersonal relationships. For a college student, failing in the campus social life may negatively affect the student academically, socially, and mentally, which in turn may reduce her/him overall satisfaction at the university. For instance, ineffective peer group social connections may lead to loneliness in college students, which may result in depression, low self-esteem, and poor academic performance (Blai, 1989). Moreover, these negative social behaviors may seriously damage the career path of students.

In order to measure the association of self-efficacy with social behavior, research conducted by Smith and Betz (2000) studied 354 undergraduate students participating in a psychology course at Ohio State University. Smith and Betz found high consistency in a development sample of 354 undergraduate students with respect to the Scale of Perceived Social Efficacy (PSSE) which measures the level of confidence in a variety of social situations. They discovered a robust relationship between social efficacy and the career development phase in college students (Smith and Betz, 2000).

In light of this finding, in this survey study, we will attempt to find out if self-efficiency and campus life satisfaction of METU students are associated, as well as how the social aspects correspond to academic performance and major satisfaction.

## 2.3-) Literature Review on Academic Performance

Intuitively, a strong academic background and skill set are important to college achievement. However, it is normally believed that a host of other students' personal and institutional attributes impact student attitudes or their satisfaction with the college experience. Ajzen and Fishbein (1980) theorized that an individual's intentions, and thus their behavior, may be predicted by attitudes. From this basis, other researchers have offered that student satisfaction supports their intention to stay in college, which supports student retention (Martirosyan, Saxon, & Wanjohi, 2014).

Kamemera et. al (2003) reported that student satisfaction with the learning environment and student services was correlated with their performance. Palak and Walls (2009) found a positive relationship between satisfaction and achievement. Dryden, Webster, and Fraser (2010) maintained that achievement was not related to satisfaction with learning except for students with the highest satisfaction ratings. Learning was most effective with high satisfaction, high cohesion, and low friction. The literature review showed a mixed relationship between satisfaction and academic achievement. Taking this into account, our study shifts to examining the role of students' academic performance on campus life satisfaction and investigates the relationship between the satisfaction of METU students and academic performance.

## 2.4-) Literature Review on Major Satisfaction, Social Self-Efficacy, Academic Performance & Campus Life Satisfaction

It would not be wrong to assume that there is an intuitive connection between university students' self-efficacy and major satisfaction of the university students. Komarajju, Swanson, and Nadler (2014) conducted research with 226 students to verify this hypothesis. As a result of the regression analysis, they observed that the increase in the self-efficacy of the university students boosted their course satisfaction with their major satisfaction (Komarraju, Nadler & Swanson, 2014). Likewise, we anticipate that the results of our study will demonstrate a link between self-efficacy and major satisfaction for METU students.

It is not difficult to assume that a satisfied and social student will also be successful academically. In order to prove that idea empirically, in the article *Life Satisfaction and Student Performance*, overall life satisfaction was considered as a dependent variable and the students' GPAs were examined in relation to their cumulative GPAs and life domains (Rode et al., 2005). Similarly, in our project, we are planning to examine the relationship between students' GPAs and their academic achievements by considering campus life satisfaction as a dependent variable.

### 3-) Methodology/Analysis

It was conducted a survey to students using an online platform called Jotform in order to gather information for our analysis. Then we it organized and prepared the collected data for analysis. We utilized the ggplot2 package in R-Studio to create visual representations of the data. We determined whether the variables in question were parametric or non-parametric using the Shapiro Wilk test. Based on the results, we either employed a linear regression model or an appropriate non-parametric test for further analysis.

Some of the major satisfaction questions were divided into three subgroups satisfaction with training, satisfaction with the facilities, and satisfaction with the program schedules. Responses are obtained using a five-point Likert scale ranging from 1 (not very well, strongly disagree) to 5 (very well, strongly agree). After that, subgroup responses were averaged for each person. Then, the descriptive statistics were examined. A bar graph was created to examine the means of subgroups by faculties. Shapiro Wilk normality test was applied to determine whether the satisfactions of males and females were normally distributed. In addition, the shape of the distribution was measured by calculating the kurtosis and skewness coefficients. The z-test was used to determine whether there is a significant difference between the overall major satisfaction means of females and males. The reason for this is the population variance was unknown, however, the sample size was over 30. The bar plot was established to show the trend of increasing/decreasing students' confidence in readiness for the world of work through their academic classes. Shapiro Wilk normality test and QQ-plot were applied to determine whether major satisfaction was normally distributed or not. Regression analysis was used to examine the causal relationships between academic performance, campus life satisfaction (predictor variables), and major satisfaction (response variable). A correlation chart was generated to examine the relationship between all the variables.

This survey involved the Perceived Social Self-Efficacy (PSSE) by Smith & Betz (2000). the measure consists of 25 logically formulated items that evaluate the degree of confidence in a series of social conditions. 19 of these items were relatively selected and separated into 4 subfactors: Networking (1-5) - Expressing ideas (6-9) - Teamwork(10-11) Self-Confidence and Assertiveness (12-19). The responses are collected using a five-point Likert scale ranging from 1 (not very well) to 5 (very well). In this study, Cronbach's alpha was used to determine the internal consistency reliability of the scale. The bar graphs were generated to demonstrate the distribution of social self-efficacy between two factors: Faculty and Residence. The post-Stratification sampling method by gender is applied to minimize the sampling error and potential non-response bias. The t-test was established to decide whether there is a significant association between the students with a higher social self-efficacy value with producing better academic performance.

In the academic performance part of *METU Students's Campus Life Satisfaction Survey*, the academic performance of students was evaluated based on the 6 different sub-factors. According to the average of 14 different likert questions of these factors, a scaled academic performance value was created for each student. In the following graph, the students's campus life satisfaction and their GPA relation were illustrated based on the conducted research question.

## 4-) Results and Findings

### 4.1-) Descriptive Statistics

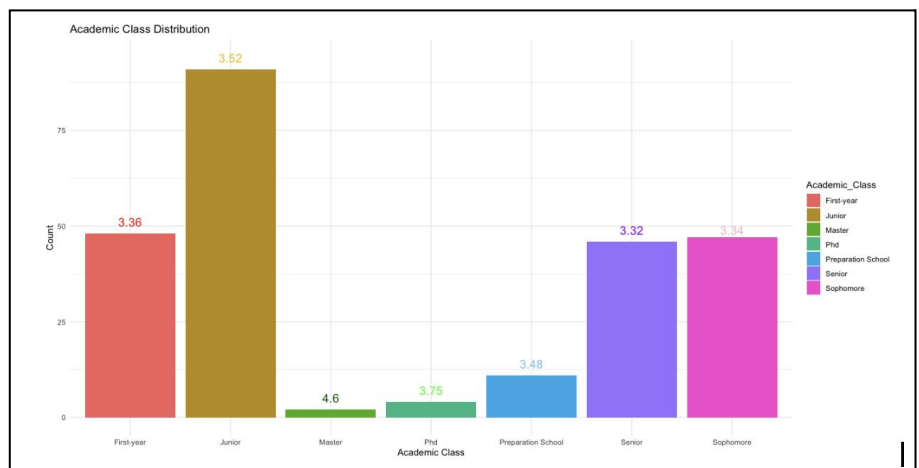

*Figure 2: Academic class distribution*

In Figure 2, their overall satisfaction scores out of five are shown. Freshman, Sophomore, and Senior counts are almost the same. The number of Junior students is the most. The numbers indicated on the graphs show the average of the satisfaction the students get from the campus out of 5. While the satisfaction received from the campus is high in prep (3.48) and junior students (3.52), these values decrease for the Senior (3,32) and Sophomore (3.34) students.

| Factors | Min | Max | Mean | StandardDeviation | Count | DistinctCount |
|---|---|---|---|---|---|---|
| Academic_Performance_Value | 1.71 | 5 | 3.334618474 | 0.675882305 | 249 | 24 |
| Campus_Life_Satisfaction_Value | 1.6 | 5 | 3.428514056 | 0.623619605 | 249 | 33 |
| Major_Satisfaction_Value | 1.2 | 5 | 3.30313253 | 0.592281526 | 249 | 44 |
| Social_Self_Efficacy_Value | 1.19 | 4.5 | 2.920080321 | 0.573631606 | 249 | 43 |

*Table 2: Satisfaction values of the factors*

In Table 2, the satisfaction values of the four factors are between 2.92 and 3.43 out of 5. The minimum value is Social Self Efficacy, and the maximum value is Campus Life Satisfaction.

```
Cronbach's alpha for the 'survey8' data-set

Items: 10
Sample units: 249
alpha: 0.79
```

Survey's Cronbach's alpha is zero point seventy-nine (see Table 3). Since this value is greater than zero point seven, the survey is consistent.

*Table 3: Cronbach's alpha for the survey*

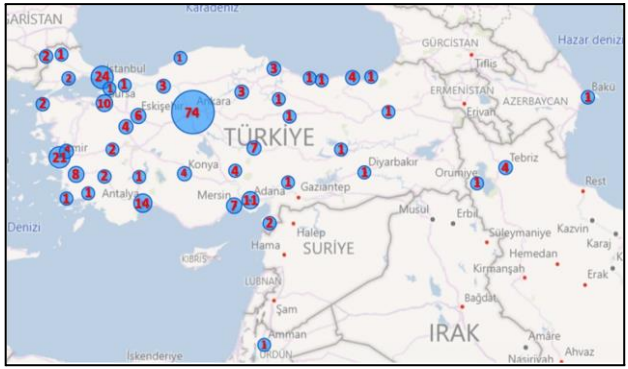

Figure 3 represents the participant's geographic distribution. The survey has a "Which city did you live in before coming to METU?" question, and this map was created according to that question. This map shows that most of the participants lived in Ankara before coming to Middle East Technical University.

*Figure 3: Participant's geographic distribution*

## 4.2-) Major Satisfaction

### 4.2.1-) Subgroups of Major Satisfaction

The subgroups seen in Table 4 are ranked according to their mean values. It has been observed that the median value of satisfaction of training is higher than the other subgroups, and the most repeated value is four. When these descriptive statistics were examined, it was concluded that METU students were more satisfied with the training.

| Subgroups | Min | Max | Mean | SD | Median | Mode |
|---|---|---|---|---|---|---|
| Training | 1.25 | 5 | 3.544177 | 0.7121188 | 3.5 | 4 |
| Facilities | 1 | 5 | 3.145582 | 0.8258918 | 3 | 3 |
| Schedules | 1 | 5 | 3.117805 | 0.8636682 | 3 | 3 |

*Table 4: Descriptive Statistics for the Major Satisfaction Subgroups (n=249)*

The means of these subgroups were investigated according to faculties. When Figure 4 is examined, it was determined that satisfaction of training was generally higher than others, and satisfaction of program schedules was less, and the results were consistent with descriptive statistics. It is observed that satisfactions in faculties are generally evenly distributed. However, when the Faculty of Economics and Administrative Sciences is examined, it is seen that the training is quite high compared to the others. On the other hand, program schedules satisfaction in the faculty of foreign languages is lower than the others. When Figure 4 is examined in general, it has been determined that the satisfaction values are mostly above three, which means neutral, and it has been concluded that METU students are satisfied with these subgroups.

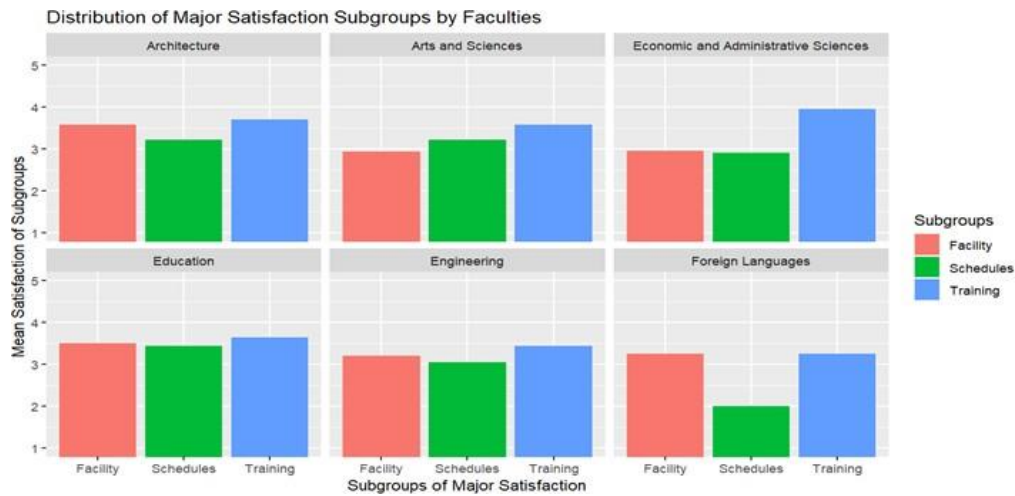

*Figure 4: Distributions of Major Subgroups by Faculties*

## 4.2.2-) Overall Major Satisfactions of Female and Male

Due to the small number of participants who chose prefer not to say and others, only the overall major satisfactions of female and male were examined. First, it was tested whether the distributions were normally distributed. According to the Shapiro Wilk normality test results in Figure 5, the null hypothesis could not be rejected because the p values for both genders were greater than significance level ($p = 0.1203 > 0.05$ for female, $p = 0.1484 > 0.05$ for male). Therefore, it has been determined that the major satisfactions of female and male are normally distributed.

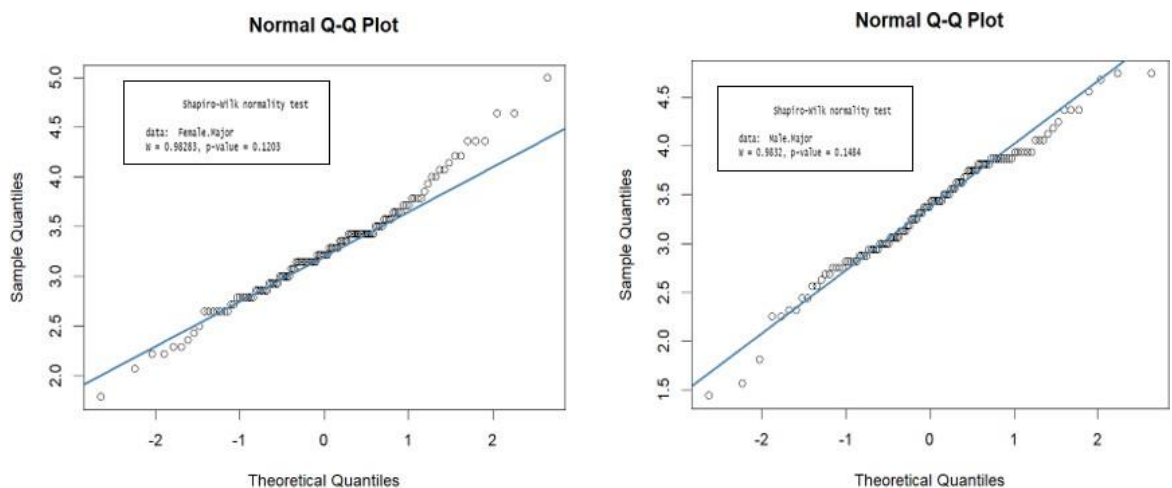

*Figure 5: Normal Q-Q Plot of Overall Major Satisfactions of Female and Male*

Since kurtosis values are greater than three for both gender (see Figure 6,7), they have leptokurtic distribution. It means that they tend to produce more outliers than the normal distribution. The skewness coefficient for female is 0.36 (see Figure 6). Because it is bigger than zero, distribution of major satisfaction of female appears to be positively skewed, and more of the values are concentrated on the left side. On the other hand, because coefficient of skewness of male (-0.30) is less than zero, distribution is negatively skewed, that is, more of the values are concentrated on the right side of the distribution (see Figure 7).

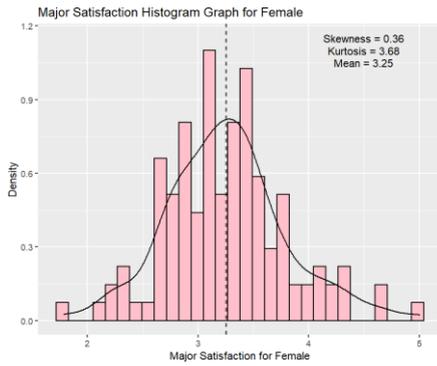 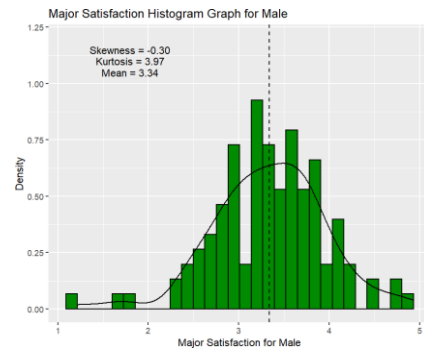

*Figure 6: Major Satisfaction of Female Histogram*          *Figure 7: Major Satisfaction of Male Histogram*

It was examined whether there was a significant difference between the overall major satisfaction means of female and male. To answer this, a hypothesis has been established. Since p value in Table 5 is bigger than significance level (0.268 > 0.05), null hyphothesis could not be rejected. This means that there is no significant difference between the mean of major satisfaction of female and mean of major satisfaction of male.

```
          Two-sample z-Test

data:  Male.Major and Female.Major
z = 1.1077, p-value = 0.268
alternative hypothesis: true difference in means is not equal to 0
95 percent confidence interval:
 -0.0637091  0.2293324
sample estimates:
mean of x mean of y
 3.337167  3.254355
```

*Table 5: R Output of Z-Test*

### 4.2.3-) Feeling Ready for The World of Work

By looking at figure 8, it can be examined that there is a significant percentage of students are unsure about their readiness for the world of work in earlier years of their education. However, as students progress through their academic classes, they become more confident in their readiness. This is indicated by a consistent decrease in the percentage of "unsure" responses and a consistent increase in the percentage of "yes" responses from the preparation school to the junior year where for the first time, the answer of "yes" is in the majority with 43.96 percent. In the senior year, a decrease is observed in the percentage of "yes" responses, and the answers to "yes" and "unsure" are equalized at 39.13 percent. In the master's year, there are no "unsure" responses. In the PhD year, again the answer of unsure is in the majority with 50 percent.

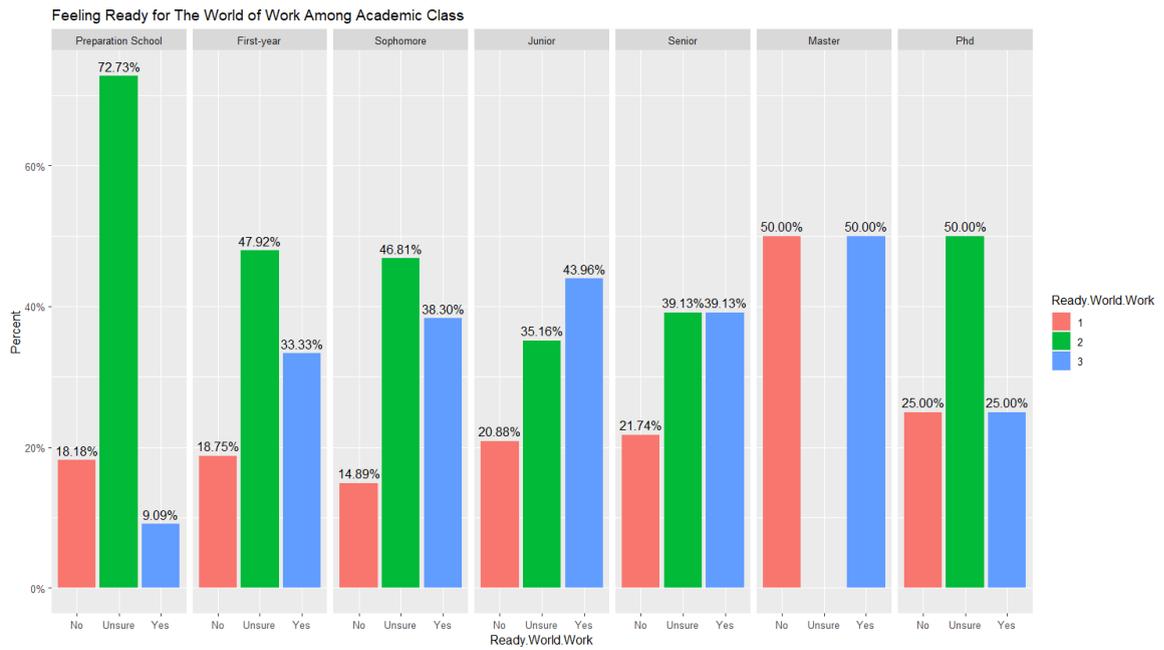

*Figure 8: Feeling Ready for The World of Work Among Academic Class*

## 4.2.4-) Regression Analysis between Academic Performance, Campus Life Satisfaction and Major Satisfaction

In this part, regression analysis was conducted between academic performance, campus life satisfaction (predictor variables), and major satisfaction (response variable). It is important to check whether the response variable (major satisfaction) is normally distributed before conducting a regression analysis.

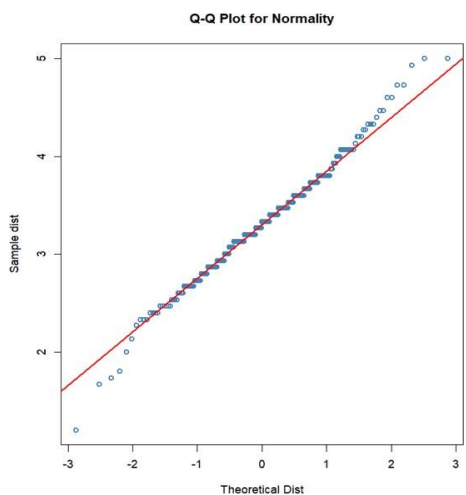

A Q-Q plot (Quantile-Quantile plot) is one way to visually check for normality. In figure 9, the points fall approximately along the Q-Q line, then the data can be assumed to be normally distributed.

```
         Shapiro-Wilk normality test
data:  survey10$Major_Satisfaction_Value
W = 0.98998, p-value = 0.08376
```

*Table 6 : Shapiro-Wilk test*

*Figure 9: A Q-Q plot for Normality*

The Shapiro-Wilk test is another way to check for normality. By looking at table 6, it can be examined that since the p-value of major satisfaction is greater than 0.05 (significance level) which is 0.08, the major satisfaciton is normally distributed. So, there is no need to do any transformation before conducting the model.

By looking at table 7, it can be examined that the adj-R square value of the model which is 0.46 indicates that these two predictor variables explain about 46% of the variation in Major Satisfaction. The F-statistic has a p-value (< 2.2e-16) less than the significance level (0.05), indicating that at least one of the independent variables in the model is significantly related to Major Satisfaction. Academic Performance and Campus Life Satisfaction p-values are (3.66e-10) and (<2e-16) respectively. Since The p-values for both predictor variables are less than 0.05, they are statistically significant.

Major Satisfaction =0.84+0.28 (Academic Performance)+0.44 (Campus Life Satisfaction)

In regression equation: The intercept 0.8447 ($\beta_0$) represents the predicted value of the major satisfaction when all predictor variables academic performance and campus life satisfaction are equal to zero.

$\beta_1 = 0.28$, indicates that while all other variables are zero, a one-unit increase in academic performance will lead to an increase of 0.28 units in Major Satisfaction.

$\beta_2 = 0.44$, indicates that while all other variables are zero, a one-unit increase in Campus Life Satisfaction will lead to an increase of 0.44 units in Major Satisfaction.

```
Call:
lm(formula = Major.Satisfaction ~ Academic.Performance + Campus.Life.Satisfaction)

Residuals:
     Min       1Q   Median       3Q      Max
-1.73327 -0.25213  0.01083  0.27431  1.63958

Coefficients:
                         Estimate Std. Error t value Pr(>|t|)
(Intercept)               0.84477    0.16754   5.042 8.94e-07 ***
Academic.Performance      0.28683    0.04390   6.534 3.66e-10 ***
Campus.Life.Satisfaction  0.44032    0.04935   8.922  < 2e-16 ***
---
Signif. codes:  0 '***' 0.001 '**' 0.01 '*' 0.05 '.' 0.1 ' ' 1

Residual standard error: 0.4314 on 246 degrees of freedom
Multiple R-squared:  0.4737,    Adjusted R-squared:  0.4694
F-statistic: 110.7 on 2 and 246 DF,  p-value: < 2.2e-16
```

*Table 7: Summary of The Model*

## 4.2.5-) Correlation Analysis between Academic Performance, Campus Life Satisfaction and Major Satisfaction

The correlation coefficients in the chart provide a summary of the relationship between the variables and can help to identify any patterns or associations between the variables. By looking at figure 10, it can be examined that the correlation between major satisfaction and campus life satisfaction (0.62) is positive which indicates that as campus life satisfaction increases, major satisfaction is likely to increase as well. The correlation is considered "strong" since a value of correlation coefficient in the interval of 0.6 to 0.799, which implies a large association between the two variables.

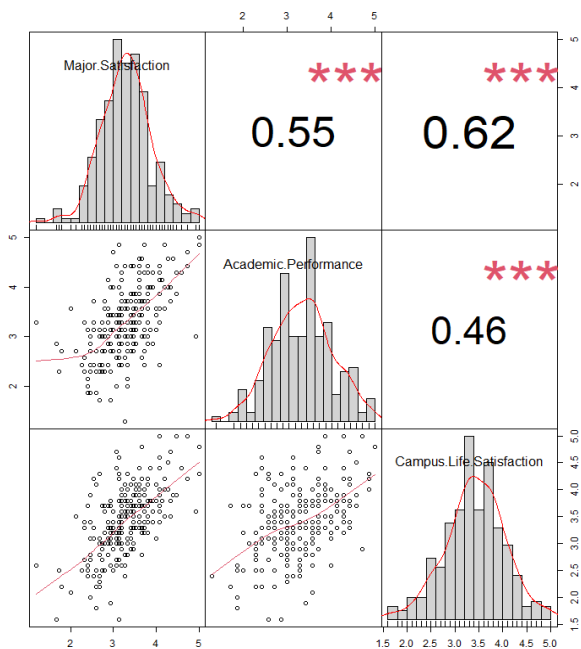

In addition, the correlation between major satisfaction and academic performance (0.55) is positive. The correlation is considered "moderate" since a value of correlation coefficient in the interval of 0.4 to 0.599. Lastly, the correlation between academic performance and campus life satisfaction (0.46) is also positive. The correlation is considered "moderate" since a value of correlation coefficient in the interval of 0.4 to 0.599.

*Figure 10: Correlation Chart*

## 4.3-) Social Self-Efficacy

### 4.3.1-) Internal Consistency Reliability Estimate for The Scale

```
Cronbach's alpha for the 'survey6' data-set

Items: 19
Sample units: 249
alpha: 0.787
```

*Table 8: Cronbach's alpha for (PSSE)*

Internal consistency reliability coefficient of 0.94 was reported by Smith and Betz (2000). Cronbach's alpha was obtained to establish the consistency of the scale. The internal consistency reliability estimate for the scale was (α = 0.78) which indicates it had acceptable consistency.

**4.3.2-) Social Self-efficacy Distribution Among Faculty and Residence**

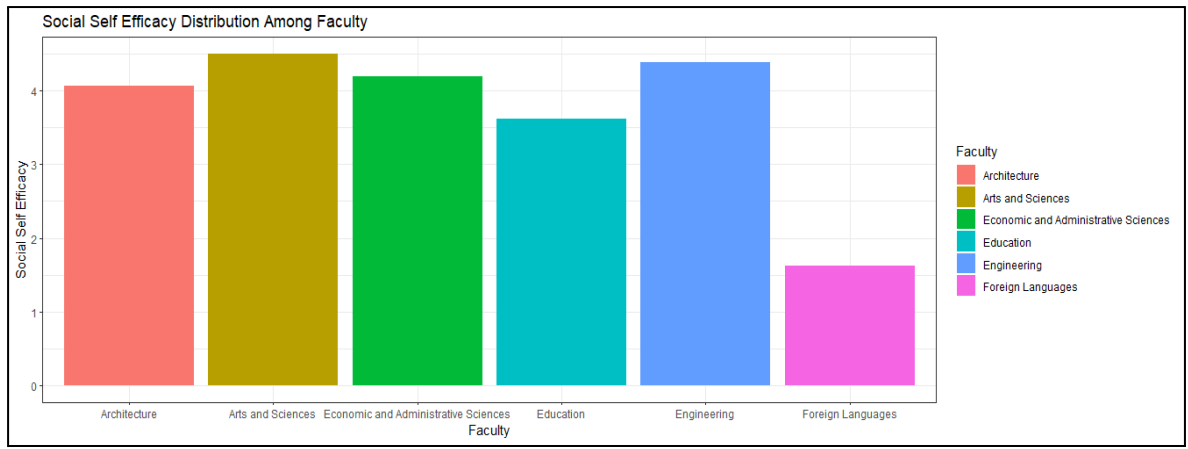

*Figure 11: Social Self-Efficacy Distribution Among Faculty*

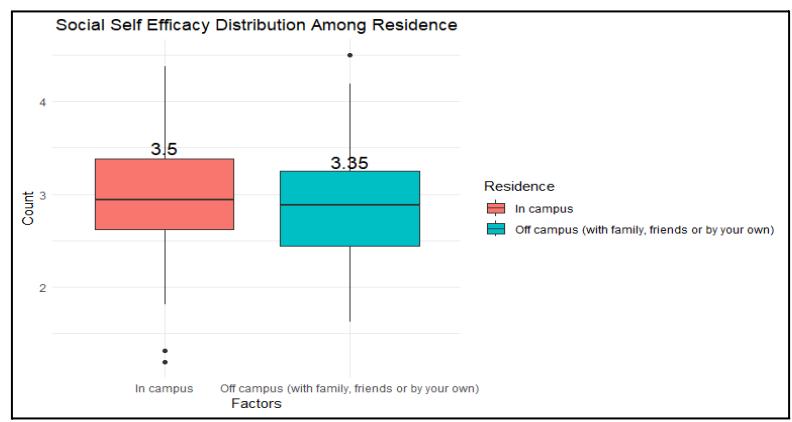

*Figure 12: Social Self-Efficacy Distribution Among Residence*

Social Self-efficacy Distribution was examined among residents and faculties to investigate their associations. As it is represented in Figure 11 Arts and Science faculty has the highest and the Foreign Languages faculty has the lowest social self-efficacy values in METU. However, it can be observed that there is no significant difference between the faculties. By looking at Figure 12 it is shown that the mean values of in-campus and off-campus residents are extremely close to each other by 3.5 and 3.35 respectively, so it is conducted that there is no relationship between Social Self Efficacy and both residents and faculties.

## 4.3.3-) Post-Stratification on Social Self-Efficacy by Gender

|  | Mean | SE |
|---|---|---|
| Social self-efficacy value | 2.9191 | 0.0366 |
|  | Total | SE |
|  | 77214 | 968.95 |

*Table 9: Gender Wise Mean and Total Value*

| Gender | Social self-efficacy value | SE |
|---|---|---|
| Female | 2.934634 | 0.05038247 |
| Male | 2.906949 | 0.05206446 |

*Table 10: Post-Stratification on Social Self-Efficacy by Gender*

With Post-Stratification on Social Self-Efficacy by Gender in our data, from Table 9 and Table 10, we can examine that the overall social self-efficacy of Males and Females in METU is approximately equal to each other. Although, based on our data we can see that Females seem to have a slight (1.03%) more social self-efficacy effect. At the same time, the small standard error (~0.05), shows a very small variability in our data. Males have a higher variance, so their variability is bigger in METU. Both genders have a mean of lower than 3 so both groups seem not to have a well social self-efficacy. Even after calculating the 95% confidence interval, we can still observe that both upper and lower bounds are lower than 3. And the total mean (2.91) is also lower than 3.

## 4.3.4-) Hypothesis Test of Social Self Efficacy In Relation to Academic Performance

```
        Welch Two Sample t-test

data:  social4_higher$Academic_Performance_Value and social4_lower$Academic_Performance_Value
t = 4.0466, df = 57.861, p-value = 0.000156
alternative hypothesis: true difference in means is not equal to 0
95 percent confidence interval:
 0.2307754 0.6826269
sample estimates:
mean of x mean of y
 3.714286  3.257585
```

*Table 11: Output of t-test*

In investigating whether Students with higher Social Self Efficacy significantly produce better academic performance a hypothesis has been established. Since the P value in Figure 4 is less than the significance level (0.000156 < 0.05) we reject the null hypothesis. As a result, students with a 3.5 or higher Social Self Efficacy Value significantly produce better academic performance than students with a 3.5 and lower Social Self Efficacy Value.

## 4.4-) Academic Performance

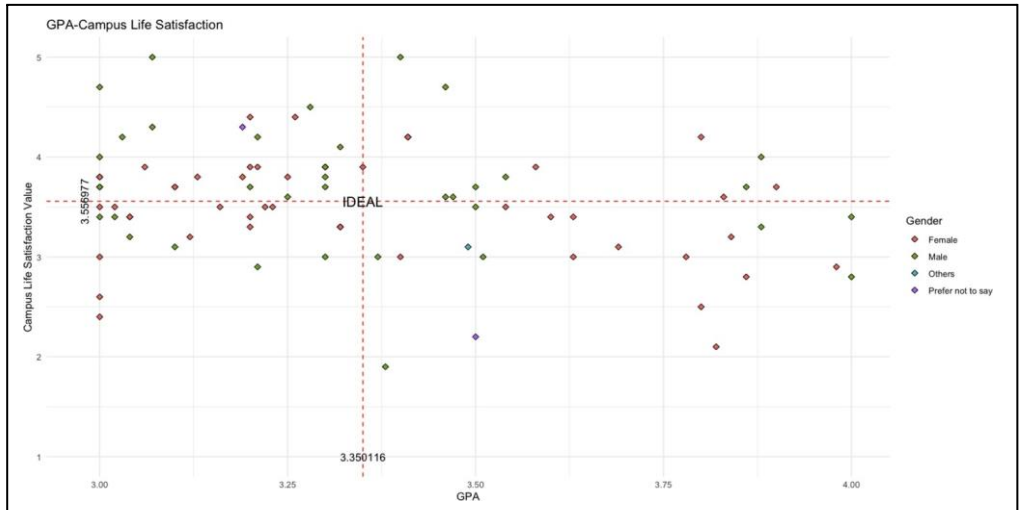

*Figure 13: GPA relationship with campus life satisfaction*

In this graph, in which students with a GPA of 3 and above are compared with the overall satisfaction from campus life, the average campus life satisfaction value is 3.55 and the corresponding average GPA is 3.35. Moreover, It was investigated whether there is a difference between the campus satisfaction of university students whose average is above 3 and below 3. Welch T test was used to test the stated hypotheses because the variance values were different. The p value was found to be 0.00769 at the 95 percent confidence interval. For this reason, it was observed that there was a significant difference between the campus satisfaction of the compared student groups. Test results can be examined in detail in the following table.

```
        Welch Two Sample t-test

data:  gpa.under3$Campus.Life.Satisfaction.Value and gpa.above3$Campus.Life.Satisfaction.Value
t = -2.6947, df = 185.83, p-value = 0.00769
alternative hypothesis: true difference in means is not equal to 0
95 percent confidence interval:
 -0.39287144 -0.06076459
sample estimates:
mean of x mean of y
 3.330159  3.556977
```

*Table 12: T-test output*

### 4.4.1-) Student Attendance in Classes and Recitations

In the following figure, class attendance and recitation attendendance are illustrated as a percentage.

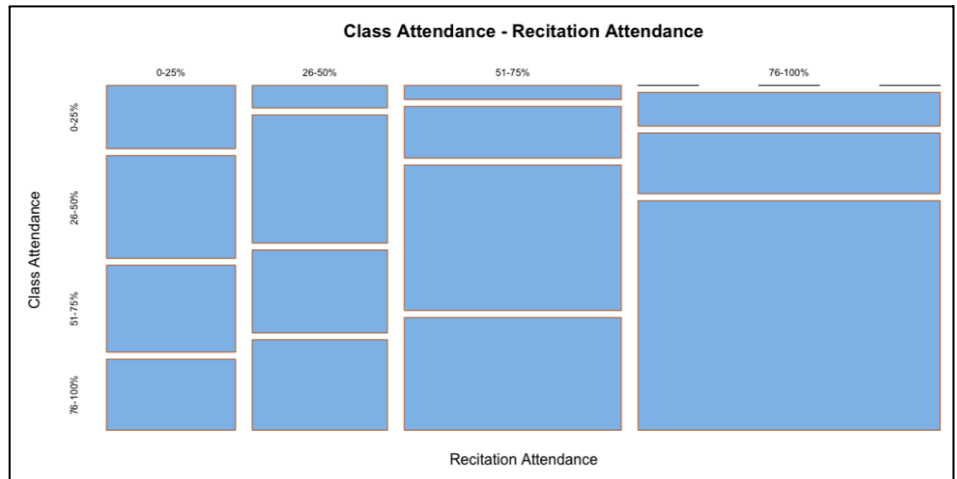

*Figure 14: Class attendance and recitation attendance*

As a minor research topic, the effect of class participation and recitation participation on GPA was desired to be observed in a descriptive manner. For this reason, the group with the highest class and recitation participation was selected and the distribution of this group was visualized in the figure given below.

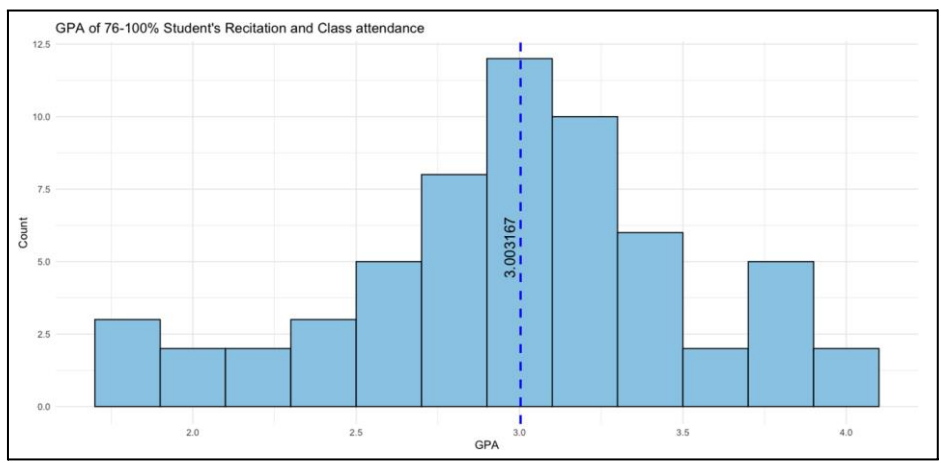

*Figure 15: GPA of 75-100% student's recitation and class attendance*

### 4.4.2-) Regression Analysis for Academic Performance and Campus Life Satisfaction

By looking at *Figure 15,* it can be examined that the R square value of the model which is 0.28 indicates the academic performance value explain about 28% of the variation in campus life satisfaction value. The F-statistic has a p-value (< 1.27e-05) less than the significance level (0.05), indicating that academic performance value is significantly related to campus life satisfaction value.

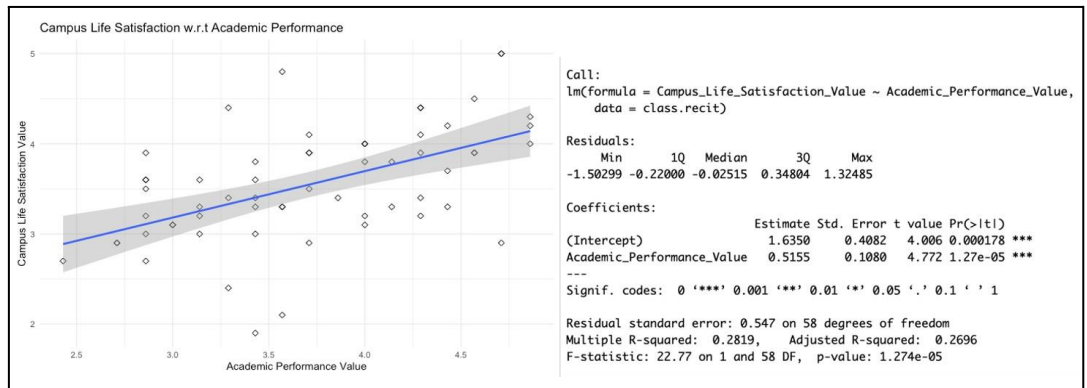

*Table 13: Summary of The Model*

## 4.5-) Campus Life Satisfaction

The graph of the last question, in which students were asked to estimate the most important factor affecting campus life, is given below.

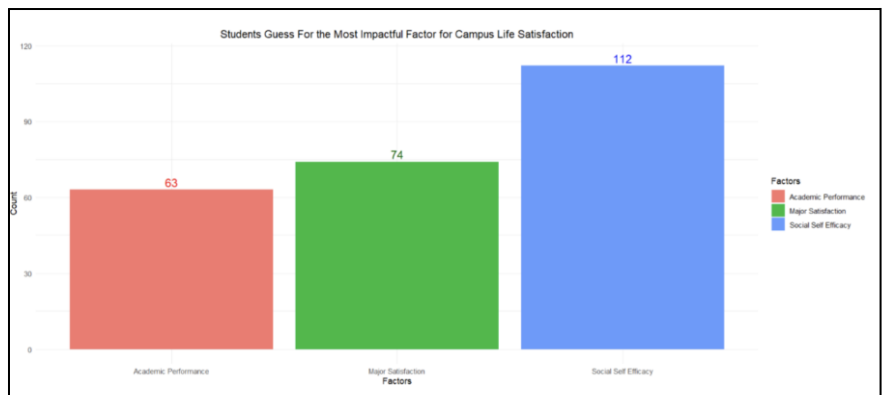

*Figure 16: Students' guess for the most important factor affecting campus life*

As seen in the graph, the factor that the students thought to have the most impact on campus life was social self efficacy.

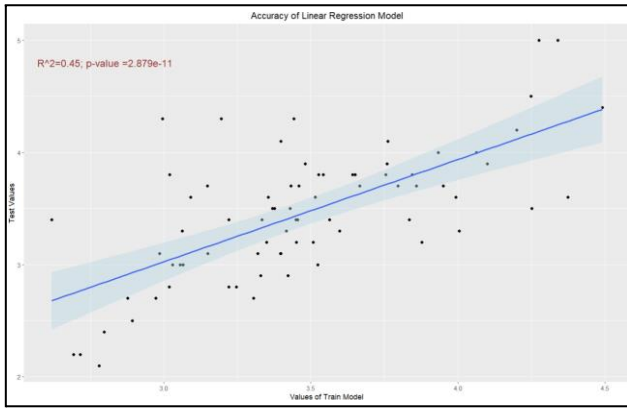 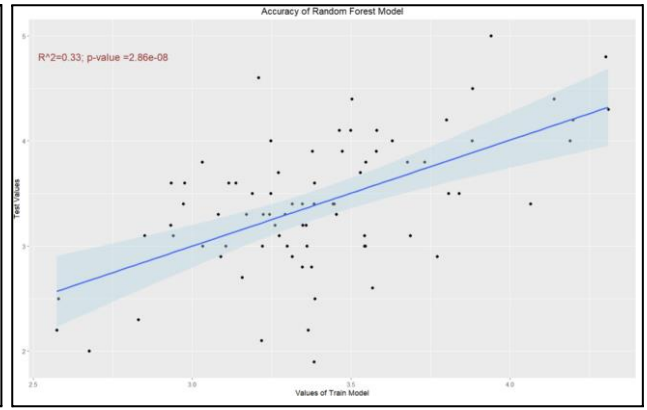

*Figure 17: Accuracy of linear regression model*     *Figure 18: Accuracy of random forest model*

Modeling of student satisfaction with major satisfaction, social efficacy and academic performance is shown in the graphs above. According to the Linear Regression model, academic performance has 0.01 p values, social self efficacy has 0.002 p values, and major satisfaction has 2e-16 p values. According to the random forest model, social self efficacy %IncMSE value was 7, academic performance %IncMSE value was 19 and major satisfaction %IncMSE value was 22. Although the order of influence of the factors affecting the campus life in which the two models are compared is different, it has been determined that major satisfaction is the most effective factor in both models.

## 5-) Discussion/Conclusion

Academic performance, social self-efficacy and major satisfaction are three important factors that effect the students's overall satisfaction. For example, high academic performance can lead to a sense of accomplishment and pride, which can positively impact a student's life satisfaction. Furthermore, good academic performance can open doors to various opportunities such as scholarships, internships, and employment opportunities. Beside academic performance of students, social self-efficacy can lead such as a sense of belonging, positive relationships with peers and faculty, and access to resources and support can all contribute to a student's satisfaction with their college experience. Additionally, a positive campus environment, including the physical surroundings and overall atmosphere, can also play a role in a student's satisfaction. However, most of all, the student's satisfaction with the department is most important factor that effects the overall satisfaction of students. Research has shown that students who are satisfied with their major tend to have better academic performance. Additionally, major satisfaction can also influence a student's decision to pursue a graduate degree or enter a specific field of work after graduation. For this reason, the first thing that students should pay attention to when choosing a university is major satisfaction.

## 6-) References

# METU STUDENTS' COLLEGE LIFE SATISFACTION SURVEY

**Factors Affecting On METU Students' Life Satisfaction**

*We are 3rd-year Statistics Students. We are conducting this survey to determine which factors impact the students' satisfaction with college life. We would like to know your social level, satisfaction with your major, and academic performance under the course responsibility. Please complete this approximately 15-minutes short survey to let us know how satisfied you are with your overall student experience. Your responses are anonymous, and none of the information will be shared with any third party, so feel free to provide honest feedback. Thank you for your participation.*

---

## Demographic Questions

This part of the survey gathers demographic (age, gender...) data

**1-) Which gender do you identify as? ***

○ Female
○ Male
○ Others
○ Prefer not to say

**2-) What is your age? ***

[                                    ]

**3-) What is your academic class standing? ***

○ Preparation School
○ First-year
○ Sophomore
○ Junior
○ Senior
○ Master
○ Phd

**4-) In what Faculty are you registered? Select all that apply. ***

☐ Architecture                    ☐ Arts and Sciences
☐ Economic and Administrative Sciences    ☐ Education
☐ Engineering                    ☐ Foreign Languages

**5-) What is your Major in METU? ***

[dropdown]

**6-) What is your student residency? ***

○ In campus
○ Off campus (with family, friends or by your own)

**7-) Which year did you first start studying at METU? ***

[text field]

Example: 2019

**8-) Which city did you live in before coming to METU? ***

[text field]

## Major Satisfaction

**9-) How likely do you think your high school advisor had effect on your preference of METU? ***

            1   2   3   4   5

Not very likely  ○  ○  ○  ○  ○  Very likely

**10-) How satisfied are you with the facilities?**

| | Very Dissatisfied | Dissatisfied | Neutral | Satisfied | Very Satisfied |
|---|---|---|---|---|---|
| **Classrooms capacity** | ○ | ○ | ○ | ○ | ○ |
| **Physical conditions of the classrooms** | ○ | ○ | ○ | ○ | ○ |
| **Lecture materials** | ○ | ○ | ○ | ○ | ○ |
| **Equipment of the department's computer lab** | ○ | ○ | ○ | ○ | ○ |

**11-) How satisfied are you with the program schedules?**

|  | Very Dissatisfied | Dissatisfied | Neutral | Satisfied | Very Satisfied |
|---|---|---|---|---|---|
| Course schedule | ○ | ○ | ○ | ○ | ○ |
| Major's courses' attendance policy | ○ | ○ | ○ | ○ | ○ |
| Breaks between courses | ○ | ○ | ○ | ○ | ○ |

**12-) How would you rate the content of the department courses? ***

           1  2  3  4  5

Not very well  ○ ○ ○ ○ ○  Very well

**13-) How would you rate the way you were assessed was a fair test of your skills? ***

           1  2  3  4  5

Not very well  ○ ○ ○ ○ ○  Very well

**14-) How would you rate the academic staff of the department is sufficient in terms of education and training? ***

           1  2  3  4  5

Not very well  ○ ○ ○ ○ ○  Very well

**15-) How would you rate the instructors' knowledge of the principles learning? ***

           1  2  3  4  5

Not very well  ○ ○ ○ ○ ○  Very well

**16-) Do you think that your training make you ready for the world of work? ***

○ Yes
○ No
○ Unsure

**17-) Do you think that you will feel competent for the job after graduating from this department? ***

○ Yes
○ No
○ Unsure

**18-) How likely are you to continue attending this department next year? ***

           1  2  3  4  5

Not very likely  ○ ○ ○ ○ ○  Very likely

**19-) How likely are you to recommend this department to others?** *

 1 2 3 4 5
Not very likely ○ ○ ○ ○ ○ Very likely

**20-) How would you rate your over all satisfaction with your Major?** *

 1 2 3 4 5
Not very well ○ ○ ○ ○ ○ Very well

# Social Self Efficacy

**21-) How well do you become friends with other people?** *

 1 2 3 4 5
Not Very Well ○ ○ ○ ○ ○ Very Well

**22-) How well do you stay friends with other people?** *

 1 2 3 4 5
Not Very Well ○ ○ ○ ○ ○ Very Well

**23-) How well would you make friends in a group where everyone else knows each other?** *

 1 2 3 4 5
Not Very Well ○ ○ ○ ○ ○ Very Well

**24-) How well do you help someone you have recently met become a part of the group to which you belong?** *

 1 2 3 4 5
Not Very Well ○ ○ ○ ○ ○ Very Well

**25-) How well do you start a conversation with a person you do not know very well?** *

 1 2 3 4 5
Not Very Well ○ ○ ○ ○ ○ Very Well

**26-) How well do you express your opinion to people who are talking about something of interest**

to you? *

                    1  2  3  4  5

Not Very Well  ○ ○ ○ ○ ○  Very Well

**27-) How well do you keep your side of your opinion even though your friends disagree with you? ***

                    1  2  3  4  5

Not Very Well  ○ ○ ○ ○ ○  Very Well

**28-) How well would you share with a group of people an interesting experience you once had? ***

                    1  2  3  4  5

Not Very Well  ○ ○ ○ ○ ○  Very Well

**29-) How well would you tell other people that they are doing something that makes you uncomfortable? ***

                    1  2  3  4  5

Not Very Well  ○ ○ ○ ○ ○  Very Well

**30-) How well do you work in harmony with other people? ***

                    1  2  3  4  5

Not Very Well  ○ ○ ○ ○ ○  Very Well

**31-) How well do you ask someone for help when you need it? ***

                    1  2  3  4  5

Not Very Well  ○ ○ ○ ○ ○  Very Well

**32-) Do you go to parties where you don't know anyone? ***

○ Yes
○ No

**33-) Do you volunteer to lead a group or organization? ***

○ Yes
○ No

**34.a-) How well would you participate in the conversation?** *

                     1   2   3   4   5

Not Very Well  ◯ ◯ ◯ ◯ ◯  Very Well

**35.a-) How well would you control those feelings?** *

                     1   2   3   4   5

Not Very Well  ◯ ◯ ◯ ◯ ◯  Very Well

**36.a-) How well did you participate in group activities?** *

                     1   2   3   4   5

Not Very Well  ◯ ◯ ◯ ◯ ◯  Very Well

## Academic Performance

**37-) What is your current CGPA?** *

[                    ]

(If you are in your first semester in first grade or prep school then you can write 0)

**38-) What percentage of the classes have you attended so far?** *

◯ 0-25%
◯ 26-50%
◯ 51-75%
◯ 76-100%

**39-) What percentage of the recitations have you attended so far?** *

◯ 0-25%
◯ 26-50%
◯ 51-75%
◯ 76-100%

**40-) How important do you think the tasks assigned to you by your professor are?** *

                         1   2   3   4   5

Very unimportant  ◯ ◯ ◯ ◯ ◯  Very important

**41-) Have you ever participated in any research at METU?** *

○ Yes
○ No

**42-) Have you ever participated in a scientific study organized by an institution or organization?** *

○ Yes
○ No

**43-) Do you have a scientific article published in an academic journal?** *

○ Yes
○ No

**44-) Have you received certified training that will contribute to your education outside of your university?** *

○ Yes
○ No

**45-) How many course(s) have you failed?** *

○ 0
○ 1
○ 2
○ 3
○ 4+

**46-) How would you rate your interaction with your professors IN the classroom?** *

             1   2   3   4   5
Very Poor ○ ○ ○ ○ ○ Very Good

**47-) How would you rate your interaction with your professors OUT classroom?** *

             1   2   3   4   5
Very Poor ○ ○ ○ ○ ○ Very Good

**48-) How well do you schedule your time?** *

             1   2   3   4   5
Very Poor ○ ○ ○ ○ ○ Very Good

**49-) How do you evaluate your preparation for each lecture by reviewing your notes?** *

|   | 1 | 2 | 3 | 4 | 5 |   |
|---|---|---|---|---|---|---|
|   | ○ | ○ | ○ | ○ | ○ |   |

**50-) How would you evaluate your preparation for exams?** *

|  | 1 | 2 | 3 | 4 | 5 |  |
|---|---|---|---|---|---|---|
| Very Poor | ○ | ○ | ○ | ○ | ○ | Very Good |

# Students' College Life Satisfaction

**51-) What do you think about extracurricular (non academic) activities on Campus?** *
☐ 1. They are too few
☐ 2. They are just right
☐ 3. I was overwhelmed with the number

**52-) Please choose the extracurricular(non-academic) activity that satisfies you the most.** *
[dropdown]

**53-) How would you evaluate the below statements in terms of student life?**

|  | Very Dissatisfied | Dissatisfied | Neutral | Satisfied | Very Satisfied |
|---|---|---|---|---|---|
| Get training/skill in a special field | ○ | ○ | ○ | ○ | ○ |
| Satisfy self needs | ○ | ○ | ○ | ○ | ○ |
| Gain knowledge about the world | ○ | ○ | ○ | ○ | ○ |

**54-) How would you evaluate your campus life satisfaction in terms of the below statements?**

|  | Very Dissatisfied | Dissatisfied | Neutral | Satisfied | Very Satisfied |
|---|---|---|---|---|---|
| Campus Location | ○ | ○ | ○ | ○ | ○ |
| Facilities in the Campus | ○ | ○ | ○ | ○ | ○ |
| Safety in Campus | ○ | ○ | ○ | ○ | ○ |
| On-campus expenses | ○ | ○ | ○ | ○ | ○ |
| Course variability | ○ | ○ | ○ | ○ | ○ |

**55-) How much do you feel you belong to METU?** *

                 1  2  3  4  5

Not Very Well  ◯ ◯ ◯ ◯ ◯  Very Well

**56-) Please rate your overall satisfaction with the university on a scale of 1 to 5.** *

                 1  2  3  4  5

Not Very Well  ◯ ◯ ◯ ◯ ◯  Very Well

*We have one last question for you.*

**57-) What do you think would have the most impact on the Students' college life satisfaction?** *

◯ Major Satisfaction
◯ Social Self Efficacy
◯ Academic Performance

Submit